\documentclass[oldversion]{aa}
\usepackage{epsfig}
\usepackage{natbib}
\usepackage{lscape}
\usepackage{wasysym}
\usepackage{array}
\usepackage{amssymb}
\usepackage{subfigure}
\usepackage{longtable}
\usepackage{rotating}
\usepackage{color}
\usepackage{colortbl}
\usepackage{soul}
\usepackage[normalem]{ulem}
\usepackage{fancyheadings}
\bibpunct{(}{)}{;}{a}{}{,}

\begin{document}

\title{Metallicity of M dwarfs} 
\subtitle{II. A comparative study of photometric metallicity
  scales\thanks{Based on observations collected with the FEROS
  spectrograph at la Silla observatory under ESO programs
  073.D-0802(A), 074.D-0670(A), 078.D-0760(A),  and with the
  ELODIE and SOPHIE spectrographs at the Observatoire de Haute Provence.
   }}

\author{ V. Neves\inst{1,2,3} \and X. Bonfils\inst{2} \and
  N. C. Santos\inst{1,3} \and X. Delfosse\inst{2} \and
  T. Forveille\inst{2}  \and F. Allard\inst{4}  \and
  C. Nat\'ario\inst{5,6} \and C. S. Fernandes\inst{5} \and
  S. Udry\inst{7}}

\institute{
Centro de Astrof{\'\i}sica, Universidade do Porto, Rua das Estrelas,
4150-762 Porto, Portugal \\
email: {\tt vasco.neves@astro.ua.pt}
\and
UJF-Grenoble 1 / CNRS-INSU, Institut de Plan\' etologie et
d'Astrophysique de Grenoble (IPAG) UMR 5274, Grenoble, F-38041,
France.
\and
Departamento de F\'{\i}sica e Astronomia, 
Faculdade de Ci\^{e}ncias, Universidade do Porto, 
Rua do Campo Alegre, 4169-007 Porto, Portugal
\and
Centre de Recherche Astrophysique de Lyon, UMR 5574: CNRS,
Universit\'e de Lyon, \'Ecole Normale Sup\'erieure de Lyon, 46 All\'ee
d'Italie, F-69364 Lyon Cedex 07, France
\and
Centro de Astronomia e Astrof\'isica da Universidade de Lisboa, Campo
Grande, Ed. C8 1749-016 Lisboa, Portugal
\and
Leiden Observatory, Leiden University, The Netherlands
\and
Observatoire de Gen\`eve, Universit\'e de Gen\`eve, 51 Chemin des
Maillettes, 1290 Sauverny, Switzerland
}


\date{Received/Accepted}

\abstract{  Stellar parameters are not easily derived from M dwarf spectra,
  which are dominated by complex bands of diatomic and triatomic 
  molecules and do not agree well with the individual line predictions of 
  atmospheric models. M~dwarf metallicities are therefore most
  commonly derived through less direct techniques. Several recent
  publications propose calibrations that provide the metallicity
  of an M~dwarf from its $K_{s}$~band absolute magnitude and its $V-K_{s}$
  color, but disagree at the $\pm$0.1 dex level.
  We compared these calibrations using a sample of 23 M dwarfs, which we
  selected as wide ($>$ 5 arcsec) companions of F-, G-, or K-
  dwarfs with metallicities measured on a homogeneous scale and 
  which we require to have $V$ band photometry measured to better than 
  $\sim$0.03~magnitude. 
  We find that the Schlaufman \& Laughlin (2010, A\&A, 519, A105+) calibration has 
  the lowest offsets and residuals against our sample, and used our 
  improved statistics to marginally refine that calibration. With
  more strictly selected photometry than in previous studies, the 
  dispersion around the calibration is well in excess 
  of the [Fe/H] and photometric uncertainties. This suggests that 
  the origin of the remaining dispersion is astrophysical rather 
  than observational.


	    }

\authorrunning{Neves et al.}
\titlerunning{Metallicity of M dwarfs. II.}
\maketitle

\section{Introduction}


M dwarfs are the smallest and coldest stars of the main sequence. 
Long lived and ubiquitous, M dwarfs are of interest in many
astrophysical contexts, from stellar evolution to the structure of
our Galaxy. Most recently, interest in M dwarfs has been increased further 
by planet search programs. Planets induce higher reflex 
velocities and deeper transits when they orbit and transit M dwarfs 
rather than larger FGK stars, and the habitable zone of the less
luminous M~dwarfs are closer in. Lower mass, smaller, and possibly 
habitable planets are therefore easier to find around M~dwarfs, and 
are indeed detected at an increasing pace \cite[e.g.][]{Udry-2007b, Mayor-2009}.

Interesting statistical correlations between the characteristics of
exoplanets and the properties of their host stars have emerged from 
the growing sample of exoplanetary systems
\citep[e.g.][Bonfils et al. 2011 in prep.]{Endl-2006,Johnson-2007,Udry-2007}.
Of those, the planet-metallicity correlation was first identified 
and remains the best established: a higher metal content increases, on 
average, the probability that a star hosts Jovian planets 
\citep{Gonzalez-1997, Santos-2001a, Santos-2004b,
  Fischer-2005}. Within the core-accretion paradigm for planetary 
formation, that correlation reflects the higher mass of solid material
available to form protoplanetary cores in the protoplanetary disks of 
higher metallicity stars. The correlation is then expected to extend 
to, and perhaps be reinforced in, the cooler M dwarfs. To
counterbalance the lower overall mass of their protoplanetary disks, those
disks need a higher fraction of refractory material to form similar 
populations of the protoplanetary core. Whether the planet-metallicity 
correlation that seems to vanish for Neptunes and lower mass planets 
around FGK stars \citep{Sousa-2008, Bouchy-2009} persists for Neptune-mass planets around M dwarfs is still an open question.

Our derivation of the first photometric metallicity calibration for
M~dwarfs \citep{Bonfils-2005} was largely motivated by probing their 
planet-metallicity correlation, though only two M-dwarf planetary
systems were known at the time. A few planet detections later,
a Kolmogorov-Smirnov test of the metallicity distributions of M 
dwarfs {\it with} and {\it without} known planets indicated that 
they only had a $\sim11\%$ probability of being drawn from a 
single parent distribution \citep{Bonfils-2007}. With an improved metallicity 
calibration and a larger sample of M~dwarf planets,
\citet{Schlaufman-2010} lower the probability that M-dwarf 
planetary hosts have the same metallicity distribution as the
general M dwarf population to $\sim6\%$. This result is in line 
with expectations for the core accretion paradigm, but is
only significant at the $\sim$2~$\sigma$ level. Both finding planets 
around additional M dwarfs and measuring metallicity more precisely 
will help characterize this correlation and the possible lack thereof.
Here we explore the second avenue.

Measuring accurate stellar parameters from the optical spectra of M dwarfs 
unfortunately is not easy. As the abundances of diatomic and triatomic 
molecules (e.g. TiO, VO, H$_{2}$O, CO) in the photospheric layers increases 
with spectral subtype, their forest of weak lines eventually erases
the spectral continuum and makes a line-by-line spectroscopic
analysis difficult for all but the earlier M subtypes. 
\citet{Woolf-2005,Woolf-2006} measured atomic 
abundances from the high-resolution spectra of 67 K and M dwarfs 
through a classical line-by-line analysis, but had to restrict 
their work to the earliest M subtypes ($T_{eff} > 3500$ K) and 
to mostly metal-poor stars (median [Fe/H]$ =-0.89$ dex). They 
find that metallicity correlates with CaH and TiO band strengths,
but do not offer a quantitative calibration.

Although the 
recent revision of the solar oxygen abundance
\citep{Asplund-2009,Caffau-2011} has greatly improved the agreement 
between model atmosphere prediction and spectra of M~dwarfs observed 
at low-to-medium resolution \citep{Allard-2010}, many visual-to-red 
spectral features still correspond to molecular bands that are 
missing or incompletely described in the opacity databases that 
underly the atmospheric models. At high spectral resolution, many
individual molecular lines in synthetic spectra are additionally 
displaced from their actual position.
Spectral synthesis, as well, has therefore had limited success in 
analyzing M~dwarf spectra \citep[e.g.][]{Valenti-1998,Bean-2006}. In 
this context, less direct techniques have been developed to evaluate the
metal content of M dwarfs. Of those, the most successful leverage
the photometric effects of the very molecular bands that complicate
spectroscopic analyses. Increased TiO and VO abundances in metal-rich
M dwarfs shift radiative flux from the visible range, where these
species dominate the opacities, to the near infrared. For a 
fixed mass, an increased metallicity also reduces the bolometric 
luminosity. Those two effects of metallicity work together in the 
visible, but, in the [Fe/H] and T$_{eff}$ range of interest here, they
largely cancel out in the near-infrared. As a result, the absolute $V$ magnitude 
on an M dwarf is very sensitive to its metallicity, while its near
infrared magnitudes are not \citep{Chabrier-2000, Delfosse-2000}.
Position in a color/absolute magnitude diagram that combines visible
and near-infrared bands is therefore a sensitive metallicity probe,
but one that needs external calibration.

We pioneered that approach in \citet{Bonfils-2005}, where we anchored
the relation on a combination spectroscopic metallicities of early-M dwarfs
from \citet{Woolf-2005}  and metallicities, which we measured for
the FGK primaries of binary systems containing a widely separated
M~dwarf component. That calibration, in terms of the $K_{s}$-band absolute 
magnitude and the $V-K_{s}$ color, results in a modestly significant 
disagreement between the mean metallicity of solar-neighborhood 
early/mid-M dwarfs and FGK dwarfs. \citet{Johnson-2009} correctly 
points out that M and (at least) K dwarfs have the same age 
distribution, since both live longer than the age of the universe, 
and that they are therefore expected to have identical metallicity 
distributions. They derived an alternative calibration, anchored in
FGK+M binaries that partly overlap the \citet{Bonfils-2005} sample,
which forces the agreement of the mean metallicities of local samples
of M and FGK dwarfs. Most recently, \citet{Schlaufman-2010} have pointed 
out the importance of kinematically matching the M and GK samples before
comparing their metallicity distributions, and used  stellar structure
models of M~dwarfs to guide their choice of a more effective
parametrization of position in the $M_{K_{s}}$ vs $V-K_{s}$ diagram. The
difference between the three calibrations varies slightly across the
Herzprung-Russell diagram but, on average, the \citet{Johnson-2009}
calibration is 0.2 dex more metal-rich than \citet{Bonfils-2005}, and
\citet{Schlaufman-2010} is half-way between those two extremes. Those
discrepancies are largely irrelevant when comparing M~dwarfs 
with metallicities consistently measured on any of these three scales, 
but they are uncomfortably large in any comparison with external
information.

We set out here to test those three calibrations. For that purpose,
we have assembled a sample of 23 M dwarfs with accurate photometry, 
parallaxes, and metallicity measured from a hotter companion
(Sect.~\ref{sample}). We then perform statistical tests of the
three calibrations in Sect.~\ref{test}, and in Sect.~\ref{latest}
we discuss those results and slightly refine the
\citet{Schlaufman-2010} calibration, which we find works best.
Section~\ref{discussion} presents our conclusions, and an appendix 
compares our preferred calibration against metallicities obtained
with independent techniques.

\section{Sample and observations}
\label{sample}


We adopt the now well-established route of measuring the metal 
content of the primaries of FGK$+$M binaries through classical 
spectroscopic methods, by assuming that it applies to the M 
secondaries. We searched for such binaries in the third edition
of the catalog of nearby stars \citep{Gliese-1991}, the catalog of 
nearby wide binary and multiple systems \citep{Poveda-1994}, 
the catalog of common proper-motion companions to $Hipparcos$ 
stars \citep{Gould-2004}, and the catalog of disk and halo binaries 
from the revised Luyten catalog \citep{Chaname-2004}. To ensure 
uncontaminated measurements of the fainter M secondaries, 
we required separations of at least 5~arcsec. That initial selection 
identified almost 300 binaries. We eliminated known fast rotators, 
spectroscopic binaries, pairs without a demonstrated common 
proper motion, as well as systems that do not figure in the revised 
$Hipparcos$ catalog \citep{van_Leeuwen-2007} from which we 
obtained the parallaxes of the primaries. With very few exceptions,
the secondaries have good $JHK_{s}$ photometry in the 2MASS
catalog \citep{Skrutskie-2006}, which we therefore adopt as 
our source of near-infrared photometry. The only exception
is Gl~551 (Proxima Centauri), which has saturated $K_{s}$ 2MASS measurements
and for which we use the \citet{Bessell-1991} measurements 
that we transform into $K_{s}$ photometry using the equations of \citet{Carpenter-2001}. 

Precise optical photometry of the secondaries, to our initial
surprise, has been less forthcoming, and we suspected noise in 
their $V$-band photometry to contribute much of the dispersion seen 
in previous photometric metallicity calibrations. We therefore 
applied a strict threshold in our literature search and only
retained pairs in which the $V$-band magnitude of the secondary 
is measured to better than 0.03 magnitude. This criterion 
turned out to severely restrict our sample, and we plan
to obtain $V$-band photometry for the many systems that meet
all our other requirements, including the availability of a good
high-resolution spectrum of the primary. \citet{Mermilliod-1997}
has been our main source of Johnson-Cousins $VRI$ photometry.
For ten sources $RI$ photometry was in Weistrop and Kron systems instead
of Johnson-Cousins. We therefore applied transformations following
\citet{Weistrop-1975} and \citet{Leggett-1992}, respectively. 
The $RIJH$ photometry was used to calculate metallicity from 
the \citet{Casagrande-2008} calibration, as discussed in the Appendix.
Our final sample contains 23 systems, of which 19 have M-dwarf 
secondaries and four have K7/K8 secondaries.

We either measured the metallicity of the primaries from 
high-resolution spectra or adopted measurements from the 
literature which are on the same metallicity scale. We 
obtained spectra for nine stars 
with the FEROS spectrograph \citep{Kaufer-1998} on the 2.2m ESO/MPI 
telescope at La Silla. We used the ARES program \citep{Sousa-2007} 
to automatically measure the equivalent widths of the \ion{Fe}{1} and \ion{Fe}{2} 
weak lines ($< 200$ m\AA) in the Fe line list of \citet{Sousa-2008}.
This list is comprised of 263 \ion{Fe}{1} and 36 \ion{Fe}{2} stable lines, ranging, in wavelength, from 4500 to 6890 \AA.
Then, we followed the procedure described in \citet{Santos-2004b}: 
[Fe/H] and the stellar parameters are determined by imposing 
excitation and ionization equilibrium, using the 2002 version of 
the MOOG \citep{Sneden-1973} spectral synthesis program with 
a grid of ATLAS9 plane-parallel model atmospheres \citep{Kurucz-1993}.

\begin{table*}[]
\caption[]{Stellar parameters measured from the primaries, with the [Fe/H]
  of the M dwarf secondary inferred from the primary.}
\label{parameters}
\centering
\begin{tabular} {l l c c c r c c}
\hline
\hline
Primary  & Secondary & $T_{eff}$ & log g & $\xi_{t}$ & \multicolumn{1}{c}{[Fe/H]} & [Fe/H] & $T_{eff}$ \\
              &              & [K] & [cm s$^{-2}$] & [km s$^{-1}$] &  & source & source\\
\hline


Gl53.1A  &  Gl53.1B  &  4705 $\pm$ 131  &  4.33 $\pm$ 0.26  &  0.76 $\pm$ 0.25  &  0.07 $\pm$ 0.12  &  \multicolumn{2}{c}{B05} \\
Gl56.3A  &  Gl56.3B  & 5394 $\pm$ 47  &  -  &  -   &  0.00 $\pm$ 0.10  &  COR  &  S08CAL \\
Gl81.1A  &  Gl81.1B  &  5332 $\pm$ 22  &  3.90 $\pm$ 0.03  &  0.99 $\pm$ 0.02  &  0.08 $\pm$ 0.02  &   \multicolumn{2}{c}{S08} \\
Gl100A  &  Gl100C  &  4804 $\pm$ 81  &  4.82 $\pm$ 0.24  &  1.25 $\pm$ 0.24  &  -0.28 $\pm$ 0.03  &    \multicolumn{2}{c}{New} \\
Gl105A  &  Gl105B  &  4910 $\pm$ 65  &  4.55 $\pm$ 0.14  &  0.77 $\pm$ 0.18  &  -0.19 $\pm$ 0.04  &  \multicolumn{2}{c}{New} \\
Gl140.1A &  Gl140.1B  &  4671 $\pm$ 65  &  4.31 $\pm$ 0.15  &  0.54 $\pm$ 0.31  &  -0.41 $\pm$ 0.04  &  \multicolumn{2}{c}{S08} \\
Gl157A   &  Gl157B  &  4854 $\pm$ 71  &  4.75 $\pm$ 0.19  &  1.31 $\pm$ 0.20  &  -0.16 $\pm$ 0.03  &  \multicolumn{2}{c}{New} \\
Gl173.1A  &  Gl173.1B  &  4888 $\pm$ 72  &  4.72 $\pm$ 0.16  &  0.97 $\pm$ 0.21  &  -0.34 $\pm$ 0.03  &  \multicolumn{2}{c}{New} \\
Gl211   &  Gl212  &  5293 $\pm$ 109  &  4.50 $\pm$ 0.21  &  0.79 $\pm$ 0.17  &  0.04 $\pm$ 0.11  &    \multicolumn{2}{c}{B05} \\
Gl231.1A  &  Gl231.1B  &  5951 $\pm$ 14  &  4.40 $\pm$ 0.03  &  1.19 $\pm$ 0.01  &  -0.01 $\pm$ 0.01  &  \multicolumn{2}{c}{New} \\
Gl250A    &  Gl250B  &  4670 $\pm$ 80  &  4.41 $\pm$ 0.16  &  0.70 $\pm$ 0.19  &  -0.15 $\pm$ 0.09  &    \multicolumn{2}{c}{B05} \\
Gl297.2A  &  Gl297.2B  &  6461 $\pm$ 14  &  4.65 $\pm$ 0.02  &  1.74 $\pm$ 0.01  &  0.03 $\pm$ 0.05  &  \multicolumn{2}{c}{New} \\
Gl324A  &  Gl324B  &  5283 $\pm$ 59  &  4.36 $\pm$ 0.11  &  0.87 $\pm$ 0.08  &  0.32 $\pm$ 0.07  &    \multicolumn{2}{c}{B05} \\
Gl559A   &  Gl551  &  5857 $\pm$ 24  &  4.38 $\pm$ 0.04  &  1.19 $\pm$ 0.03  &  0.23 $\pm$ 0.02  &  \multicolumn{2}{c}{New} \\
Gl611A   &  Gl611B  &  5214 $\pm$ 44  &  4.71 $\pm$ 0.06  &  -  &  -0.69 $\pm$ 0.03  &  \multicolumn{2}{c}{SPO} \\
Gl653  &  Gl654  &  4723 $\pm$ 89  &  4.41 $\pm$ 0.24  &  0.52 $\pm$ 0.31  &  -0.62 $\pm$ 0.04  &  \multicolumn{2}{c}{S08} \\
Gl666A    &  Gl666B  &  5274 $\pm$ 26  &  4.47 $\pm$ 0.04  &  0.74 $\pm$ 0.05  &  -0.34 $\pm$ 0.02  &  \multicolumn{2}{c}{New} \\
Gl783.2A    &  Gl783.2B  &  5094 $\pm$ 66  &  4.31 $\pm$ 0.13  &  0.30 $\pm$ 0.19  &  -0.16 $\pm$ 0.08  &    \multicolumn{2}{c}{B05} \\
Gl797A   &  Gl797B  &  5889 $\pm$ 32  &  4.59 $\pm$ 0.06  &  1.01 $\pm$ 0.06  &  -0.07 $\pm$ 0.04  &    \multicolumn{2}{c}{B05} \\
GJ3091A    &  GJ3092B  &  4971 $\pm$ 79  &  4.48 $\pm$ 0.15  &  0.81 $\pm$ 0.22  &  0.02 $\pm$ 0.04  &  \multicolumn{2}{c}{S08} \\
GJ3194A   &  GJ3195B  & 5860 $\pm$ 47  &  -  &  -   &  0.00 $\pm$ 0.10  &  SOP  &  S08CAL \\
GJ3627A   &  GJ3628B  & 5013 $\pm$ 47 &  -  &  -   &  -0.04 $\pm$ 0.10  &  SOP  &  S08CAL \\
NLTT34353  &  NLTT34357  &  5489 $\pm$ 19  &  4.46 $\pm$ 0.03  &  0.91 $\pm$ 0.03  &  -0.18 $\pm$ 0.01  &  \multicolumn{2}{c}{New} \\

\hline

\end{tabular}

\raggedright

References. [B05] \citet{Bonfils-2005}; [COR] CCF [Fe/H] derived from
spectra of the CORALIE Spectrograph; [S08CAL] T$_{eff}$ calibration
from \citet{Sousa-2008}; [S08] \citet{Sousa-2008}; [New] This paper;
[SPO] \citet{Valenti-2005}; [SOP] CCF [Fe/H] taken from spectra of the
SOPHIE Spectrograph \citep{Bouchy-2006}.

\end{table*}

For three stars, we used spectra gathered with the CORALIE
\citep{Queloz-2000} spectrograph, on the Swiss Euler 1.2~m telescope at la Silla, and SOPHIE \citep{Bouchy-2006} spectrograph, on the Observatoire de Haute Provence 1.93~m telescope. For those three stars, we use
metallicities derived from a calibration of the equivalent
width of the cross correlation function (CCF) of their spectra 
with numerical templates \citep{Santos-2002a}. We adopted
that approach, rather than a standard spectroscopic analysis, 
because those observations were obtained with a ThAr lamp 
illuminating the second fiber of the spectrographs for highest
radial velocity precision. The contamination of the stellar 
spectra by scattered ThAr light would affect stellar parameters 
measured through a classical spectral analysis, but is absorbed
(to first order) into the calibration of the CCF equivalent width
to a metallicity. That calibration is anchored onto abundances
derived with the \citet{Santos-2004b} procedures, and has been
verified to be on the same scale to within 0.01~dex \citep{Sousa-2011}.

We adopt 10 [Fe/H] determinations from previous publications
of our group \citep{Bonfils-2005,Sousa-2008}, which also used
the \citet{Santos-2004b} methods. Finally, we take one metallicity
value from \citet{Valenti-2005}. That reference derived
its metallicities through full spectral synthesis, and 
\citet{Sousa-2008} found that they are on the same scale
as \citet{Santos-2004b}.

Table \ref{parameters} lists the adopted stellar parameters 
(effective temperature, surface gravity, micro-turbulence, and metallicity)
from high-resolution spectra of the primaries. Table \ref{estrelas} lists 
parallaxes and photometry for the
full sample, along with their respective references.
Columns 1 and 3 display the names of the 
primary and secondary stars, while columns 2 and 4 display their
respective spectral types. Column 5 lists the \textit{Hipparcos} parallaxes of the 
primaries with their associated standard errors. Columns 6 to 11
contain the  $V(RI)_{c}JHK_{s}$ photometry of the secondary and their associated
errors. Column 12 contains the bibliographic references for the 
photometry.

\begin{sidewaystable*}[]
\tiny
\caption[]{ \label{estrelas}Sample of wide binaries with an FGK primary
  and an M dwarf secondary, listing the parallaxes of the primary and photometry of the secondary, along with their respective references.}
\begin{tabular}{l c l c r r r r r r r r c}

  \hline
  \hline
Primary&Sp. Type. & Secondary& Sp. Type  &$\pi$&V&R&I&J&H&K$_{s}$&V/RI/JHK source \\
            & primary               &                  &  secondary   & [mas] & [mag] & [mag] & [mag]Ê& [mag] &Ê[mag] & [mag] & &  \\

Gl53.1A  &  K4  &  Gl53.1B  &  M3  &  48.20 $\pm$ 1.06  &  13.60 $\pm$ 0.02  &  12.48 $\pm$ 0.05  &  11.01 $\pm$ 0.05  &  9.533 $\pm$ 0.039  &  8.927 $\pm$ 0.023  &  8.673 $\pm$ 0.024  &  W93 / W93 / 2MASS \\
Gl56.3A  &  K1  &  Gl56.3B  &  K7  &  37.75 $\pm$ 0.95  &  10.70 $\pm$ 0.02  &  09.84 $\pm$ 0.03  &  9.01 $\pm$ 0.03  &  8.012 $\pm$ 0.021  &  7.369 $\pm$ 0.029  &  7.190 $\pm$ 0.020  &  B90 / B90 / 2MASS \\
Gl81.1A  &  G9  &  Gl81.1B  &  K7  &  30.44 $\pm$ 0.60  &  11.20 $\pm$ 0.01  &  10.30 $\pm$ 0.01  &  9.41 $\pm$ 0.01  &  8.413 $\pm$ 0.023  &  7.763 $\pm$ 0.021  &  7.597 $\pm$ 0.027  &  C84 / C84 / 2MASS \\
Gl100A  &  K4.5  &  Gl100C  &  M2.5  &  51.16 $\pm$ 1.33  &  12.85 $\pm$ 0.01  &  11.79 $\pm$ 0.01  &  10.43 $\pm$ 0.01  &  9.181 $\pm$ 0.027  &  8.571 $\pm$ 0.029  &  8.347 $\pm$ 0.021  &  C84 / C84 / 2MASS \\
Gl105A  &  K3  &  Gl105B  &  M4  &  139.27 $\pm$ 0.45  &  11.66 $\pm$ 0.02  &  10.45 $\pm$ 0.05  &  8.87 $\pm$ 0.05  &  7.333 $\pm$ 0.018  &  6.793 $\pm$ 0.038  &  6.574 $\pm$ 0.020  &  W93 / W93 / 2MASS \\
Gl140.1A  &  K3.5  &  Gl140.1B  &  K8  &  51.95 $\pm$ 1.16  &  10.17 $\pm$ 0.01  &  - $\pm$ -  &  - $\pm$ -  &  7.436 $\pm$ 0.023  &  6.828 $\pm$ 0.023  &  6.620 $\pm$ 0.040  &  S96 / - / 2MASS \\
Gl157A  &  K4  &  Gl157B  &  M2  &  64.40 $\pm$ 1.06  &  11.61 $\pm$ 0.03  &  - $\pm$ -  &  - $\pm$ -  &  7.773 $\pm$ 0.024  &  7.162 $\pm$ 0.033  &  6.927 $\pm$ 0.031  &  U74 / - / 2MASS \\
Gl173.1A  &  K3  &  Gl173.1B  &  M3  &  32.69 $\pm$ 1.51  &  14.19 $\pm$ 0.02  &  13.05 $\pm$ 0.05  &  11.65 $\pm$ 0.05  &  10.263 $\pm$ 0.022  &  9.715 $\pm$ 0.028  &  9.421 $\pm$ 0.024  &  W93 / W93 / 2MASS \\
Gl211  &  K1  &  Gl212  &  M0  &  81.44 $\pm$ 0.54  &  09.76 $\pm$ 0.01  &  8.81 $\pm$ 0.05  &  7.76 $\pm$ 0.05  &  6.586 $\pm$ 0.021  &  5.963 $\pm$ 0.016  &  5.759 $\pm$ 0.016  &  HIP / W93 / 2MASS \\
Gl231.1A  &  G0  &  Gl231.1B  &  M3.5  &  51.95 $\pm$ 0.40  &  13.27 $\pm$ 0.02  &  12.15 $\pm$ 0.05  &  10.62 $\pm$ 0.05  &  9.088 $\pm$ 0.023  &  8.559 $\pm$ 0.042  &  8.267 $\pm$ 0.018  &  WT81 / WT81 / 2MASS \\
Gl250A  &  K3  &  Gl250B  &  M2  &  114.94 $\pm$ 0.86  &  10.08 $\pm$ 0.01  &  09.04 $\pm$ 0.01  &  7.80 $\pm$ 0.01  &  6.579 $\pm$ 0.034  &  5.976 $\pm$ 0.055  &  5.723 $\pm$ 0.036  &  L89 / L89 / 2MASS \\
Gl297.2A  &  F6.5  &  Gl297.2B  &  M2  &  44.68 $\pm$ 0.30  &  11.80 $\pm$ 0.02  &  - $\pm$ -  &  - $\pm$ -  &  8.276 $\pm$ 0.019  &  7.672 $\pm$ 0.027  &  7.418 $\pm$ 0.016  &  R04 / - / 2MASS \\
Gl324A  &  G8  &  Gl324B  &  M4  &  81.03 $\pm$ 0.75  &  13.16 $\pm$ 0.01  &  11.94 $\pm$ 0.05  &  10.27 $\pm$ 0.05  &  8.560 $\pm$ 0.027  &  7.933 $\pm$ 0.040  &  7.666 $\pm$ 0.023  &  D88 / WT77 / 2MASS \\
Gl559A  &  G2  &  Gl551  &  M6  &  772.33 $\pm$ 2.42  &  11.05 $\pm$ 0.02  &  9.43 $\pm$ 0.03  &  7.43 $\pm$ 0.03  &  5.357 $\pm$ 0.023  &  4.835 $\pm$ 0.057  &  4.31 $\pm$ 0.03\,\,\,  &  B90 / B90 / 2MASS+B91 \\
Gl611A  &  G8  &  Gl611B  &  M4  &  68.87 $\pm$ 0.33  &  14.23 $\pm$ 0.02  &  13.00 $\pm$ 0.05  &  11.38 $\pm$ 0.05  &  9.903 $\pm$ 0.021  &  9.453 $\pm$ 0.021  &  9.159 $\pm$ 0.017  &  W96 / W96 / 2MASS \\
Gl653  &  K5  &  Gl654  &  M2  &  93.40 $\pm$ 0.94  &  10.07 $\pm$ 0.01  &  9.10 $\pm$ 0.01  &  7.95 $\pm$ 0.01  &  6.780 $\pm$ 0.029  &  6.193 $\pm$ 0.021  &  5.975 $\pm$ 0.026  &  K02 / K02 / 2MASS \\
Gl666A  &  G8  &  Gl666B  &  M0  &  113.61 $\pm$ 0.69  &  08.70 $\pm$ 0.01  &  - $\pm$ -  &  - $\pm$ -  &  7.237 $\pm$ 9.999  &  5.112 $\pm$ 0.023  &  4.856 $\pm$ 0.020  &  E79 / - / 2MASS \\
Gl783.2A  &  K1  &  Gl783.2B  &  M4  &  49.04 $\pm$ 0.65  &  14.06 $\pm$ 0.02  &  12.81 $\pm$ 0.03  &  11.20 $\pm$ 0.03  &  9.627 $\pm$ 0.018  &  9.108 $\pm$ 0.015  &  8.883 $\pm$ 0.018  &  DS92 / DS92 / 2MASS \\
Gl797A  &  G5  &  Gl797B  &  M2.5  &  47.65 $\pm$ 0.76  &  11.87 $\pm$ 0.01  &  - $\pm$ -  &  - $\pm$ -  &  8.160 $\pm$ 0.020  &  7.645 $\pm$ 0.023  &  7.416 $\pm$ 0.016  &  D82 / - / 2MASS \\
GJ3091A  &  K2  &  GJ3092B  &  M  &  33.83 $\pm$ 1.00  &  15.64 $\pm$ 0.03  &  13.81 $\pm$ 0.05  &  11.97 $\pm$ 0.05  &  11.092 $\pm$ 0.023  &  10.540 $\pm$ 0.026  &  10.266 $\pm$ 0.021  &  P82 / E76 / 2MASS \\
GJ3194A  &  G4  &  GJ3195B  &  M3  &  41.27 $\pm$ 0.58  &  12.55 $\pm$ 0.02  &  11.49 $\pm$ 0.05  &  10.15 $\pm$ 0.05  &  8.877 $\pm$ 0.021  &  8.328 $\pm$ 0.023  &  8.103 $\pm$ 0.029  &  W96 / W96 / 2MASS \\
GJ3627A  &  G5  &  GJ3628B  &  M3.5  &  38.58 $\pm$ 2.17  &  14.10 $\pm$ 0.03  &  12.88 $\pm$ 0.05  &  11.31 $\pm$ 0.05  &  9.828 $\pm$ 0.022  &  9.247 $\pm$ 0.021  &  9.015 $\pm$ 0.018  &  W88 / W88 / 2MASS \\
NLTT34353  &  G5  &  NLTT34357  &  K7  &  20.73 $\pm$ 1.05  &  12.41 $\pm$ 0.02  &  11.51 $\pm$ 0.03  &  10.59 $\pm$ 0.03  &  9.595 $\pm$ 0.026  &  8.910 $\pm$ 0.026  &  8.734 $\pm$ 0.019  &  R89 / R89 / 2MASS \\

\hline
\end{tabular}

References. [2MASS] \citet{Skrutskie-2006}; [B90] \citet{Bessell-1990}; [B91] \citet{Bessell-1991};  
[C84] \citet{Caldwell-1984}; [D82] \citet{Dahn-1982}; [D88] \citet{Dahn-1988}; 
[DS92] - \citet{Dawson-1992}; [E76] \citet{Eggen-1976}; [E79] - \citet{Eggen-1979}; 
[HIP] \citet{van_Leeuwen-2007}; [K02] \citet{Koen-2002}; [L89] \citet{Laing-1989}; 
[P82] \citet{Pesch-1982}; [R04] \citet{Reid-2004};  
[R89] \citet{Ryan-1989}; [S96] \citet{Sinachopoulos-1996}; [U74] \citet{Upgren-1974}; 
 [W88] \citet{Weis-1988}; [W93] \citet{Weis-1993}; [W96] \citet{Weis-1996}; [WT77] \citet{Weistrop-1977}; [WT81] \citet{Weistrop-1981}. 
\end{sidewaystable*}

\section{Evaluating the photometric metallicity calibrations}
\label{test}
To assess the three alternative photometric calibrations,
we evaluated the mean and the dispersion of the difference
between the spectroscopic metallicities of the primaries and
the metallicities that each calibration predicts for the 
M~dwarf components. As in previous works \citep{Schlaufman-2010,
  Rojas-Ayala-2010}, we also computed the residual mean square $RMS_p$
and the squared multiple correlation coefficient 
$R^2_{ap}$ \citep{Hocking-1976}. 

The residual mean square $RMS_{p}$ is defined as
\begin{equation}
RMS_{p} = \frac{SSE_{p}}{n-p}, \\
SSE_{p} = \sum{(y_{i,model}-y_{i})^{2}},
\label {rmsp}
\end{equation}
where $SSE_{p}$ is the sum of squared residuals for a p-term model, 
n the number of data points, and p the number of free 
parameters of the model. The squared multiple correlation coefficient R$^{2}_{ap}$ is defined as 
\begin{equation}
R^{2}_{ap} = 1-(n-1)\frac{RMS_{p}}{SST}, \\
SST = \sum{(y_i-\bar{y})^{2}}.
\label {r2ap}
\end{equation}

\noindent A low RMS$_{p}$ means that the model describes the data well, while 
R$^{2}_{ap}$ close to 1 signifies that the tested model explains 
most of the variance of the data. The R$^{2}_{ap}$ can take negative
values, when the model under test increases the variance over a null model.

We recall that $p$ should be set to the number of adjusted parameters 
when a model is adjusted, but instead is zero when a preexisting
model is evaluated against independent data. We are, somewhat
uncomfortably, in an intermediate situation, with 11, 2, and 12 
binary systems in common with the samples that define the 
calibrations of \citet{Bonfils-2005}, \citet{Johnson-2009}, and
\citet{Schlaufman-2010}, and some measurements for those systems
in common. Our sample therefore is not fully independent, and
in full rigor $p$ should take some effective value between 
zero and the number of parameters in the model. Fortunately,
that number, 2 for all three calibrations, is a small fraction
of the sample size, 23. The choice of any effective $p$ between
0 and 2 therefore has little impact on the outcome. We
present results for $p = 0$, except when adjusting an update
of the \citet{Schlaufman-2010} calibration to the full sample,
where we use $p=2$ as we should. 

We evaluate the uncertainties on the offset, dispersion, RMS$_{p}$, 
and R$^{2}_{ap}$ through bootstrap resampling. We generated
100,000 virtual samples with the size of our observed sample
by random drawing elements of our sample, with repetition. We
computed the described parameters for each virtual sample,
and used their standard deviation to estimate the uncertainties.

Table \ref{stats} displays 
the defining equations of the various calibrations, their mean offset 
for our sample, the dispersion around the mean value (rms), the residual
mean square ($RMS_{p}$), the square of the multiple
correlation coefficient ($R^{2}_{ap}$), as well as their uncertainties.
The $M_{K}$ from the B05 calibration is the absolute magnitude
calculated with the $K_{s}$ photometric magnitudes and the \textit{Hipparcos} parallaxes. The $\Delta M$ 
from the B05(2) calibration is the difference between the $V$- and the $K$-band mass-luminosity relations
of \citet{Delfosse-2000}. In the JA09 calibration, the $\Delta M_{K}$ is the difference between 
the mean value of [Fe/H] of the main sequence FGK stars from the \citet{Valenti-2005} catalog (as defined by a fifth-order polynomial
$MS = \sum{a_{i}(V-K_{s})^{i}}$, where a = \{$-$9.58933, 17.3952, $-$8.88365, 2.22598, $-$0.258854, 0.0113399\}),
and the absolute magnitude in the $K_{s}$ band. The $\Delta (V-K_{s})$ in the SL10 and `This paper' calibrations is the difference
between the observed $V-K_{s}$ color and the fifth-order polynomial function of $M_{K_{s}}$ adapted from 
the previously mentioned formula from \citet{Johnson-2009}. In this case, the coefficients of the polynomial are, in increasing 
order: (51.1413, $-$39.3756, 12.2862, $-$1.83916, 0.134266, $-$0.00382023).

\begin{figure*}
\begin{center}
\subfigure[B05 Calibration]{\includegraphics[scale=0.35]{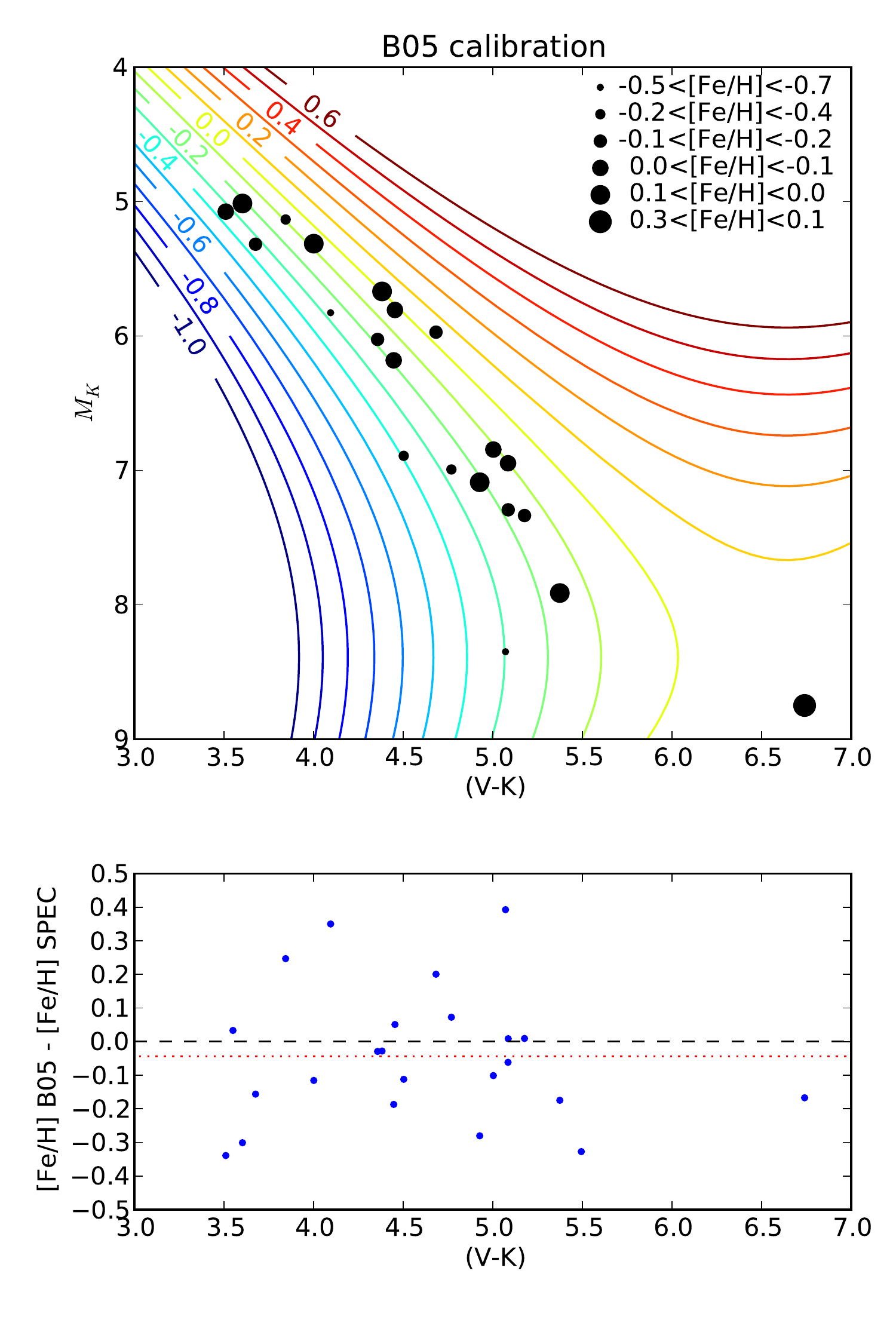}}
\subfigure[B05(2) Calibration]{\includegraphics[scale=0.35]{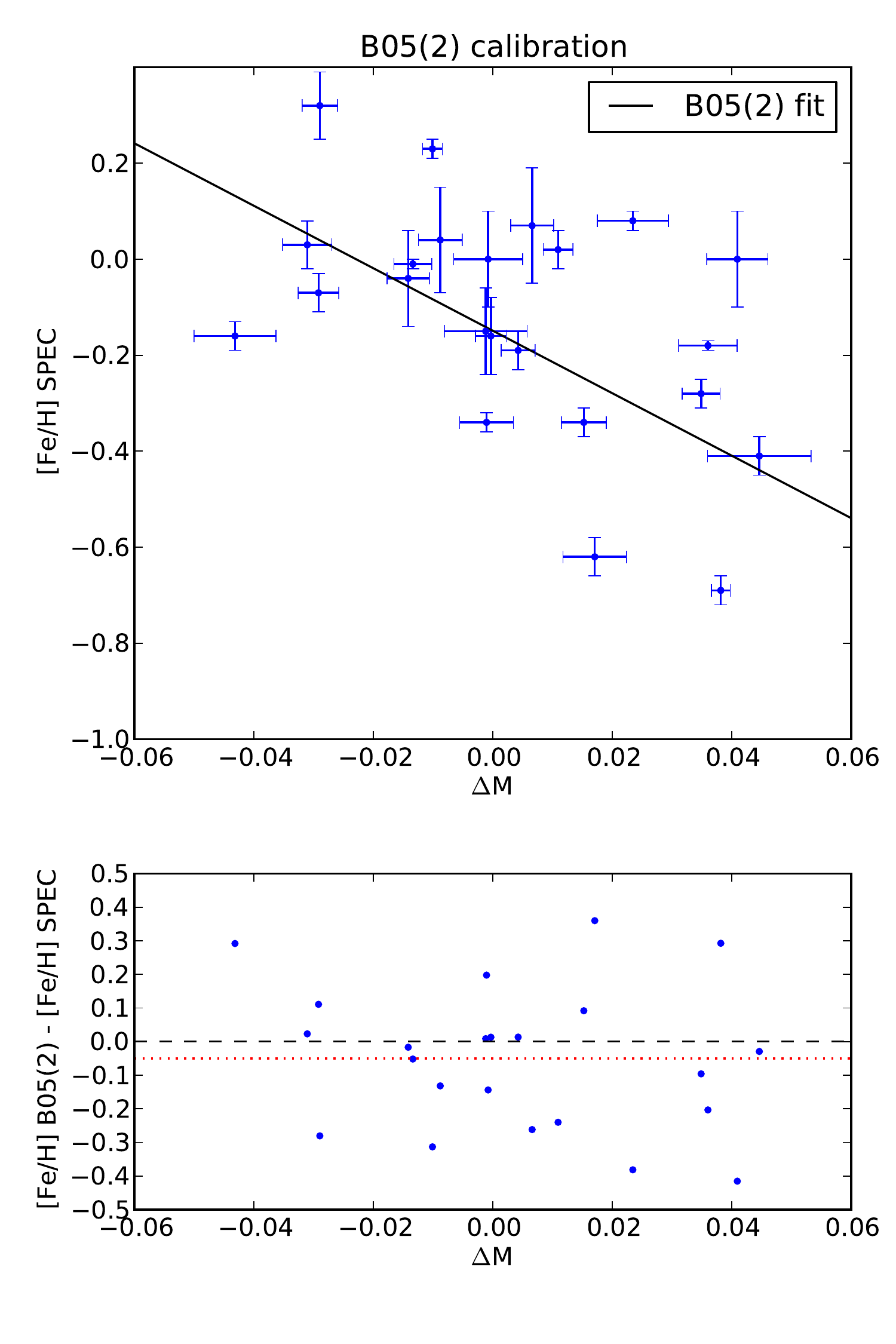}}
\subfigure[JA09 Calibration]{\includegraphics[scale=0.35]{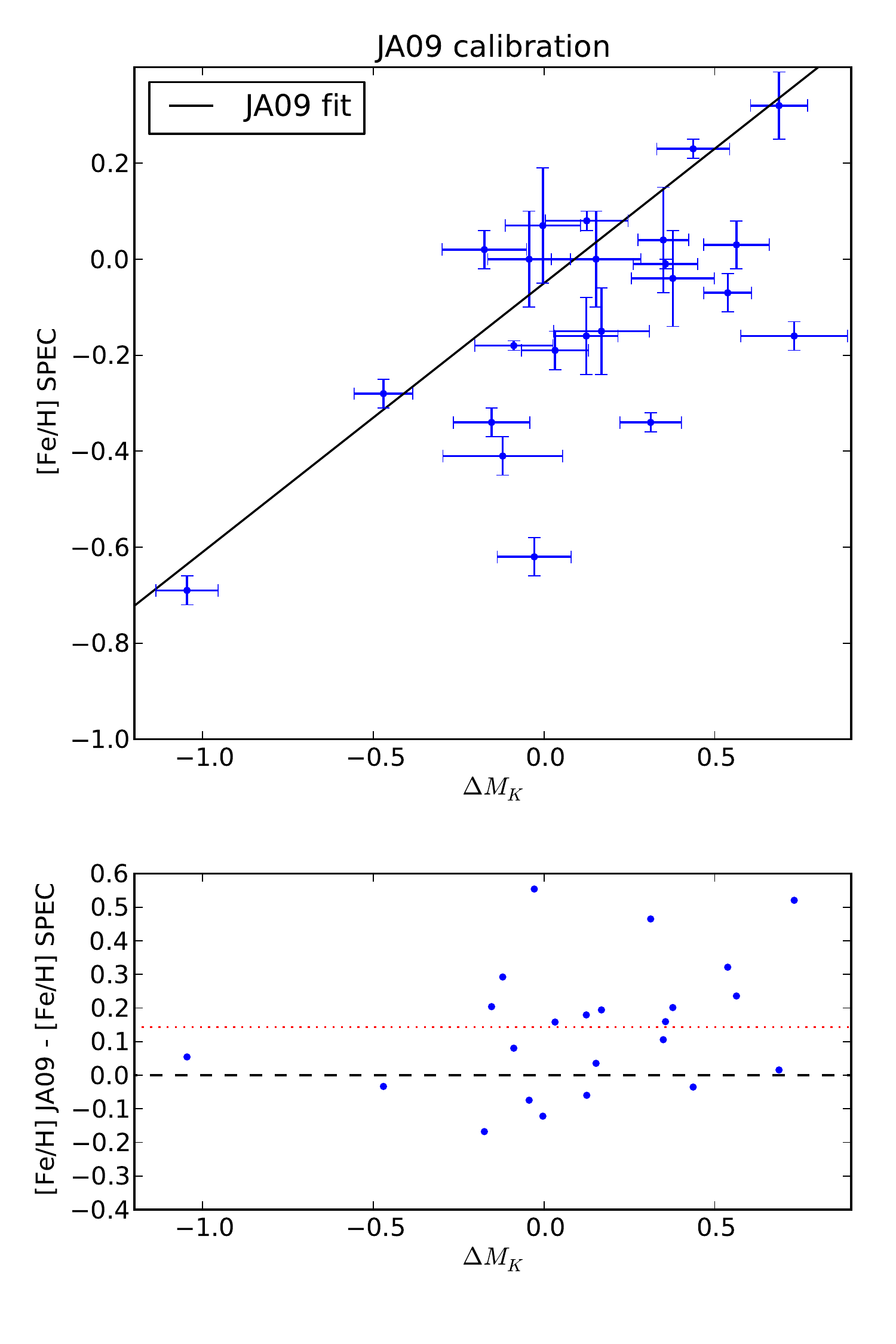}}
\subfigure[SL10 Calibration]{\includegraphics[scale=0.35]{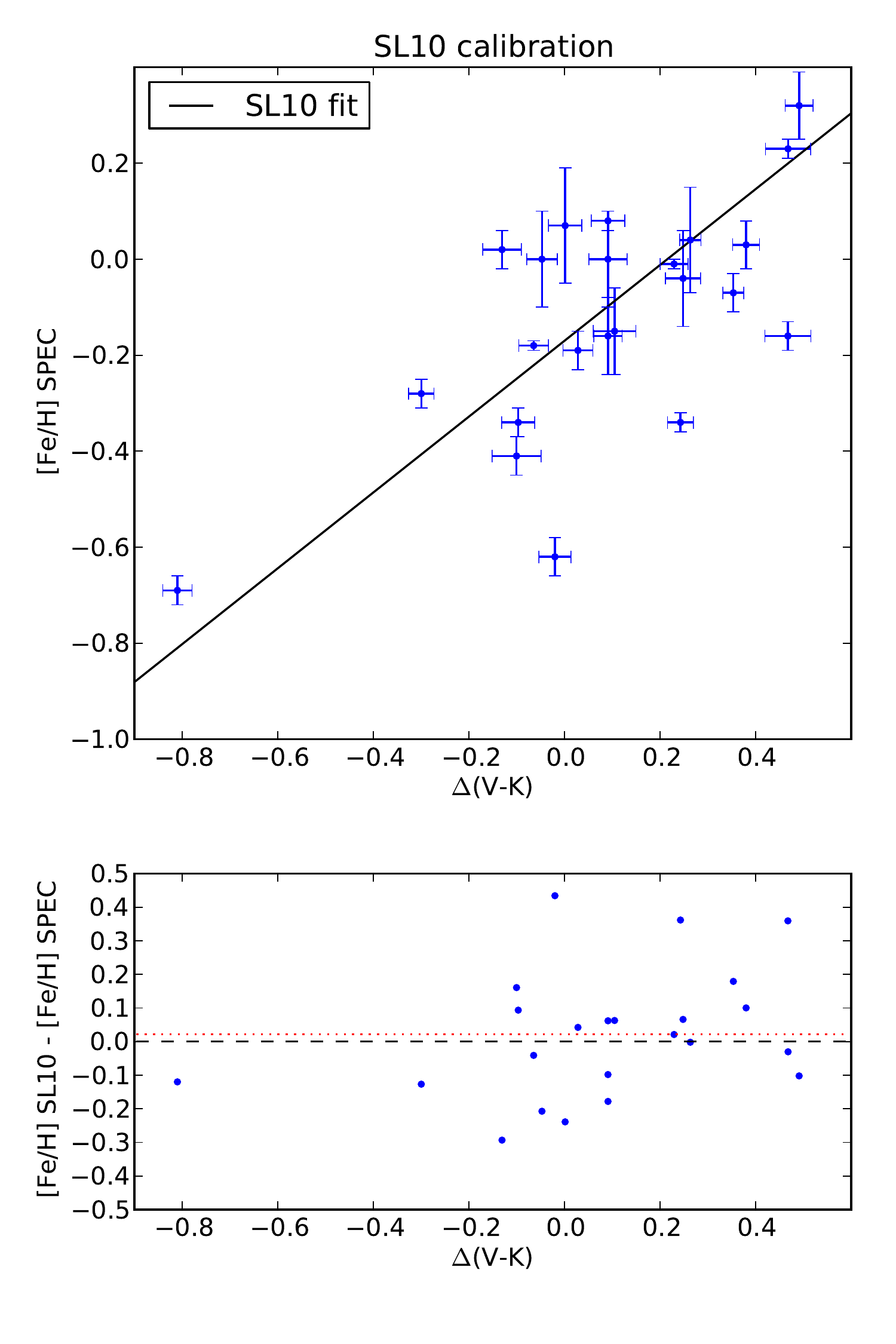}}
\subfigure[This paper]{\includegraphics[scale=0.35]{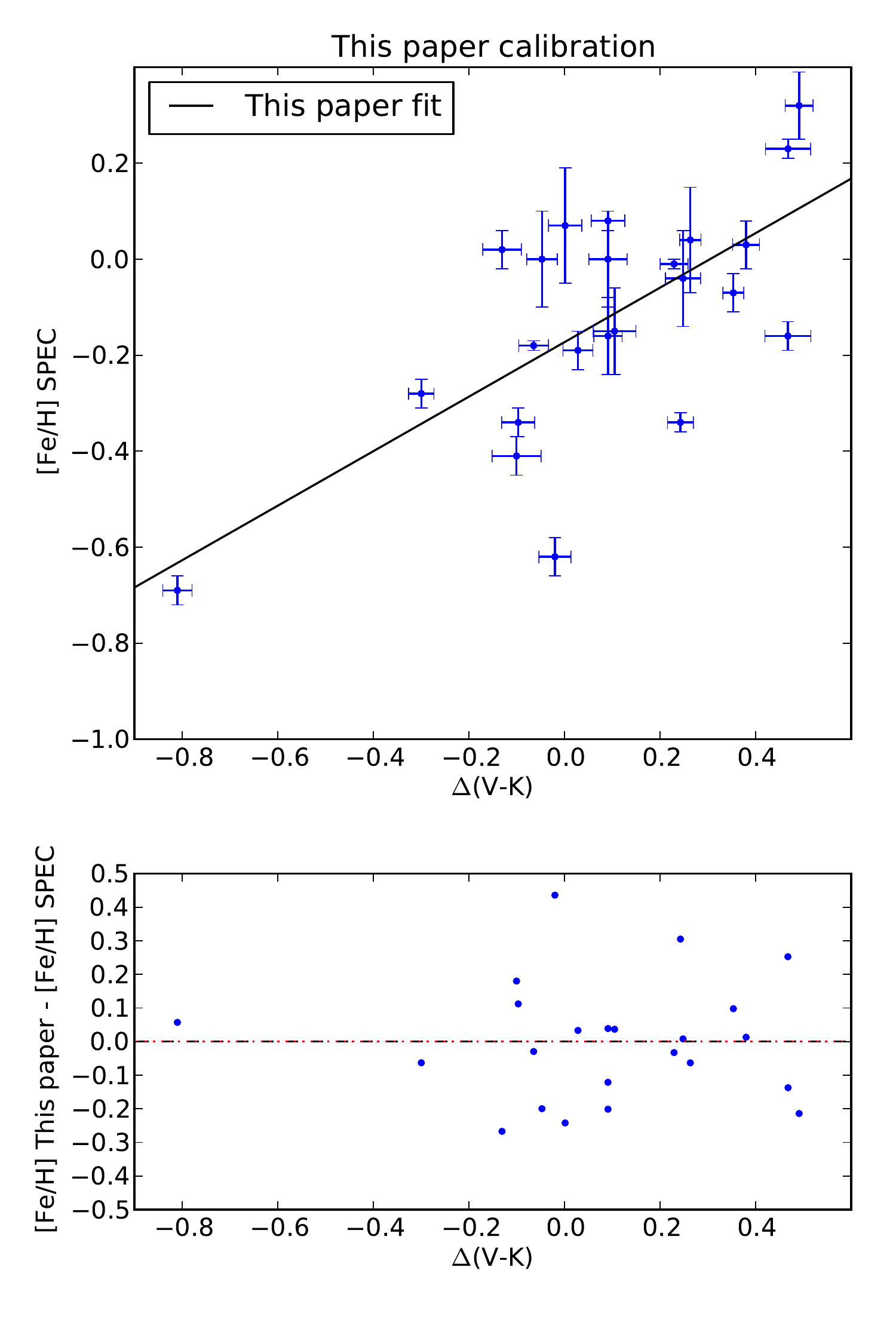}}
\end{center}
\caption{The different [Fe/H] calibrations from \citet{Bonfils-2005}
  (a and b), \citet{Johnson-2009} (c), \citet{Schlaufman-2010}
  (d), and the calibration determined in this paper (e). In each upper panel, the blue/black 
  dots represent the data points. The black line
  depicts a fit to the data except in panel (a), where the calibrated [Fe/H] is shown as isometallicity contours. 
  The lower subpanels show the difference between the calibrated and the spectroscopic metallicity. 
  The black dashed lines represent the null value,
  and the red dotted line represents the mean difference for that 
  calibration.  }

\label{fehcalib}
\end{figure*}

Figure \ref{fehcalib} depicts the different [Fe/H] calibrations from \citet{Bonfils-2005}
(a and b), \citet{Johnson-2009} (c), \citet{Schlaufman-2010}
(d), and the calibration determined in this paper (e). Table \ref{tablefehfeh} displays the metallicity values from
spectroscopy and the different calibrations, where the individual
values for each star can be compared directly.



The bootstrap uncertainties of the parameters (Table \ref{stats})
show that the rms values are the most robust. The R$^{2}_{ap}$ 
parameter, in contrast, has large uncertainties. With our small 
sample size, it therefore does not provide an effective diagnostic
of the alternative models.

\begin{table*}[]
\caption{The equations of the different calibrations, along with their calculated evaluation parameters.}
\label{stats}
\scriptsize
\begin{center}
\begin{tabular}{l r r r r}

\hline
\hline
Calibration Source + equation & offset & rms & RMS$_{P}$ & R$^{2}_{ap}$  \\
	&[dex] &[dex] & [dex] & \\ 
\hline
B05 : $[Fe/H] = 0.196 - 1.527M_{K} + 0.091M_{K}^{2} + 1.886(V-K_{s}) - 0.142(V-K_{s})^{2}$ & $-0.04\pm0.04$ & $0.20\pm0.02$ & $0.04\pm0.01$ & $0.31\pm0.22$ \\
B05(2) : $[Fe/H] = -0.149 - 6.508\Delta M, \Delta M = Mass_{V} - Mass_{K}$ & $-0.05\pm0.04$ & $0.22\pm0.02$ & $0.05\pm0.01$ & $0.21\pm0.34$ \\
JA09 : $[Fe/H] = 0.56\Delta M_{K} - 0.05, \Delta M_{K} = MS - M_{K}$ & $0.14\pm0.04$ & $0.24\pm0.04$ & $0.06\pm0.02$ & $0.03\pm0.51$ \\
SL10 : $[Fe/H] = 0.79\Delta (V-K_{s}) - 0.17, \Delta (V-K_{s}) = (V-K_{s})_{obs} - (V-K_{s})_{iso}$ & $0.02\pm0.04$ & $0.19\pm0.03$ & $0.04\pm0.01$ & $0.41\pm0.29$ \\
This paper : $[Fe/H] = 0.57\Delta (V-K_{s}) - 0.17$ & $0.00\pm0.04$ & $0.17\pm0.03$ & $0.03\pm0.01$ & $0.43\pm0.23$ \\
\hline
\end{tabular}
\end{center}
\end{table*}%

\begin{table*}[]
\caption{Spectroscopic metallicity of the primaries and 
  metallicities predicted for the secondary by the different calibrations.}
\label{tablefehfeh}
\begin{center}
\begin{tabular}{r r r r r r r r r}

\hline
\hline

Primary & Secondary & \multicolumn{5}{c}{[Fe/H] [dex]}  \\
 & & Spectroscopic & B05 & B05(2) & JA09 & SL10 & This paper \\
\hline
Gl53.1A  & Gl53.1B  & 0.07 & -0.21 & -0.19 & -0.05 & -0.17 & -0.17 \\
Gl56.3A  & Gl56.3B  & 0.00 & -0.34 & -0.42 & -0.07 & -0.21 & -0.20 \\
Gl81.1A  & Gl81.1B  & 0.08 & -0.22 & -0.30 & 0.02 & -0.10 & -0.12 \\
Gl100A  & Gl100C  & -0.28 & -0.39 & -0.38 & -0.31 & -0.41 & -0.34 \\
Gl105A  & Gl105B  & -0.19 & -0.18 & -0.18 & -0.03 & -0.15 & -0.15 \\
Gl140.1A  & Gl140.1B  & -0.41 & -0.38 & -0.44 & -0.12 & -0.25 & -0.23 \\
Gl157A  & Gl157B  & -0.16 & 0.04 & 0.13 & 0.36 & 0.20 & 0.10 \\
Gl173.1A  & Gl173.1B  & -0.34 & -0.27 & -0.25 & -0.14 & -0.25 & -0.23 \\
Gl211  & Gl212  & 0.04 & -0.08 & -0.09 & 0.15 & 0.04 & -0.02 \\
Gl231.1A  & Gl231.1B  & -0.01 & -0.11 & -0.06 & 0.15 & 0.01 & -0.04 \\
Gl250A  & Gl250B  & -0.15 & -0.18 & -0.14 & 0.04 & -0.09 & -0.11 \\
Gl297.2A  & Gl297.2B  & 0.03 & 0.00 & 0.05 & 0.27 & 0.13 & 0.05 \\
Gl324A  & Gl324B  & 0.32 & -0.01 & 0.04 & 0.34 & 0.22 & 0.11 \\
Gl559A  & Gl551  & 0.23 & 0.06 & -0.08 & 0.19 & 0.20 & 0.10 \\
Gl611A  & Gl611B  & -0.69 & -0.30 & -0.40 & -0.64 & -0.81 & -0.64 \\
Gl653  & Gl654  & -0.62 & -0.27 & -0.26 & -0.07 & -0.19 & -0.18 \\
Gl666A  & Gl666B  & -0.34 & -0.09 & -0.14 & 0.12 & 0.02 & -0.03 \\
Gl783.2A  & Gl783.2B  & -0.16 & -0.15 & -0.15 & 0.02 & -0.10 & -0.12 \\
Gl797A  & Gl797B  & -0.07 & -0.02 & 0.04 & 0.25 & 0.11 & 0.03 \\
GJ3091A  & GJ3092B  & 0.02 & -0.15 & -0.22 & -0.15 & -0.27 & -0.25 \\
GJ3194A  & GJ3195B  & 0.00 & -0.19 & -0.14 & 0.04 & -0.10 & -0.12 \\
GJ3627A  & GJ3628B  & -0.04 & -0.10 & -0.06 & 0.16 & 0.03 & -0.03 \\
NLTT34353  & NLTT34357  & -0.18 & -0.34 & -0.38 & -0.10 & -0.22 & -0.21 \\

\hline
\end {tabular}
\end{center}
\end{table*}

\section{The latest metallicity measurements and calibrations}
\label{latest}

\begin{figure*}
\begin{center}
\subfigure[B05 Calibration]{\includegraphics[scale=0.35]{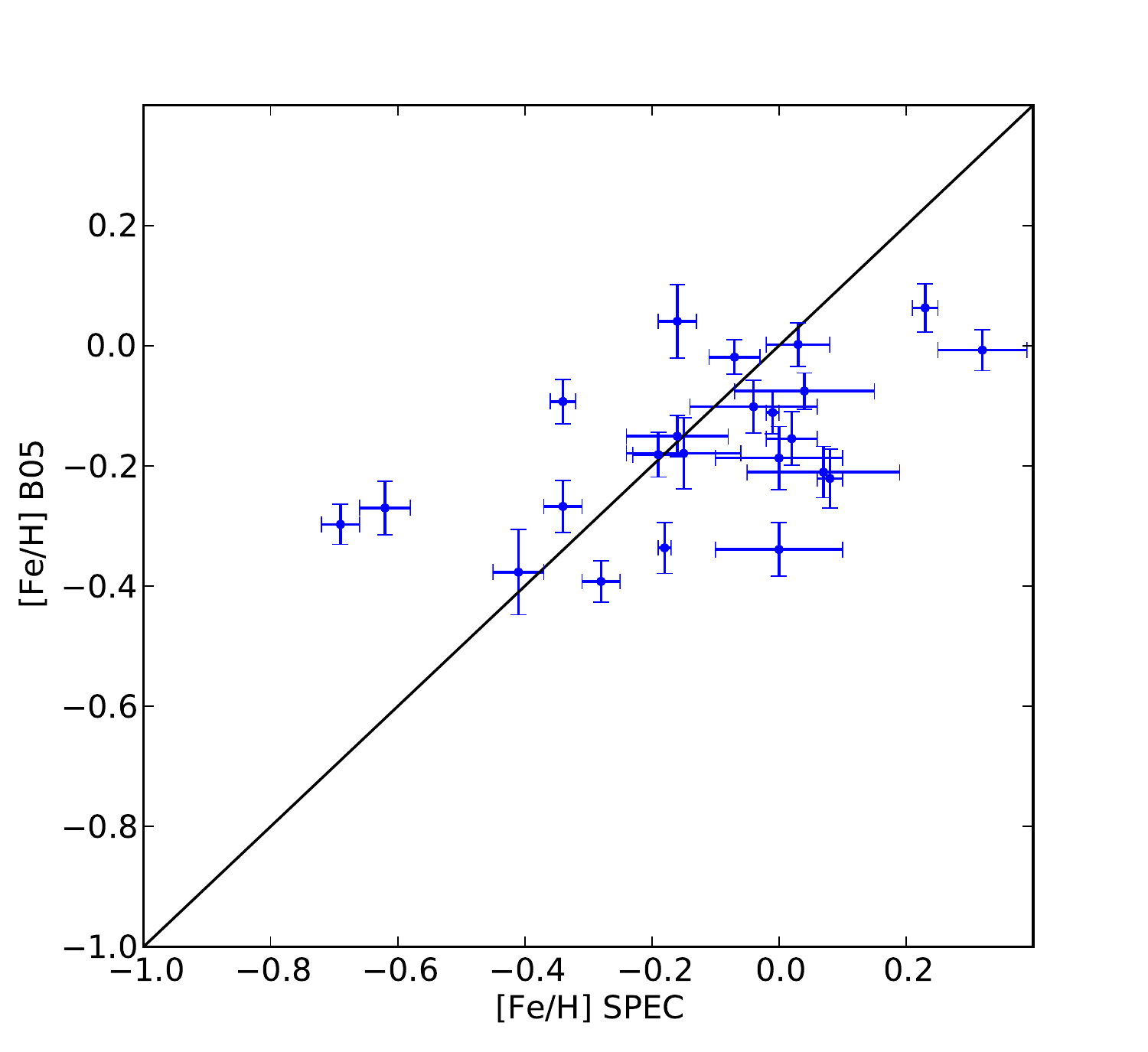}}
\subfigure[B05(2) Calibration]{\includegraphics[scale=0.35]{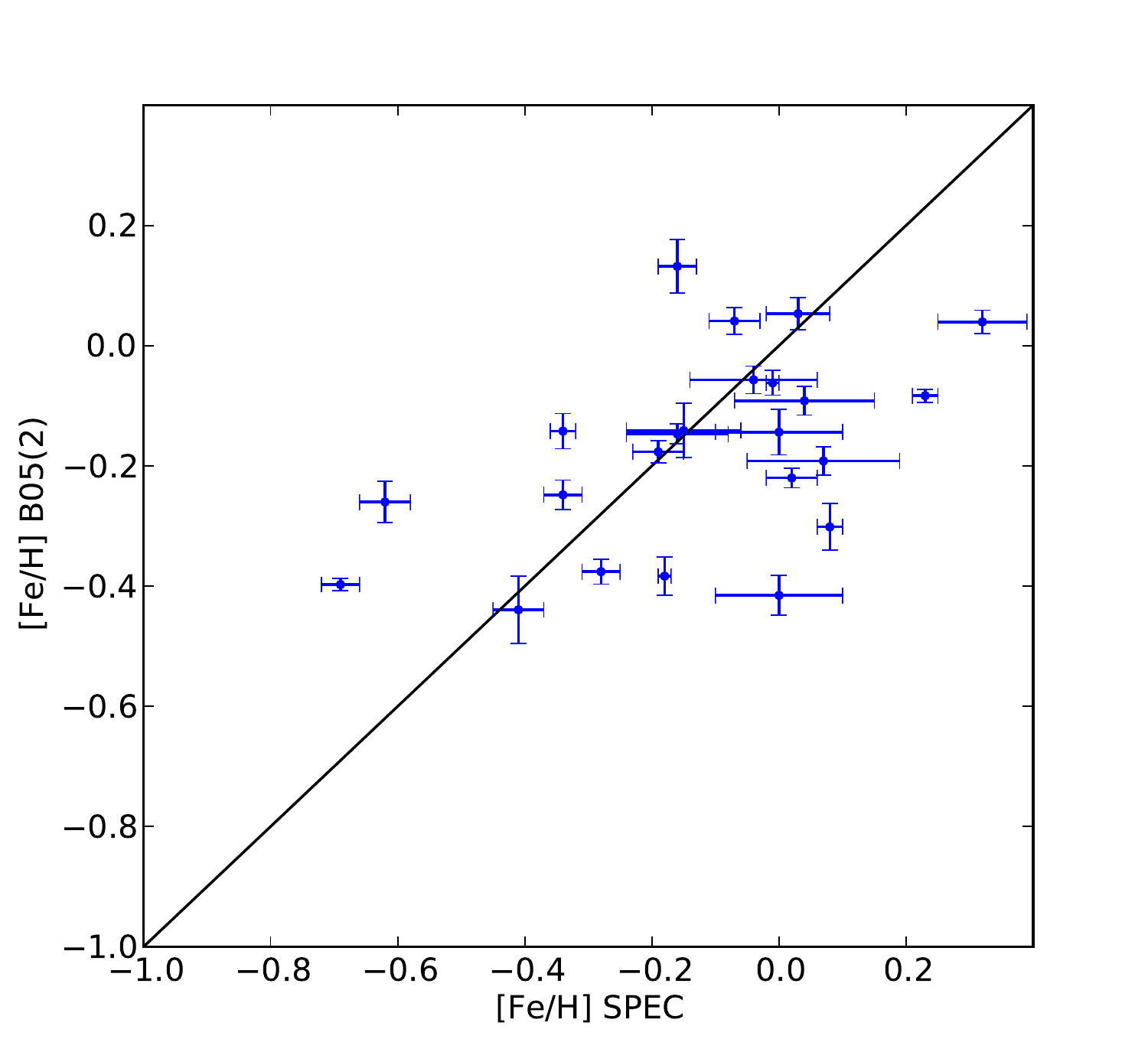}}
\subfigure[JA09 Calibration]{\includegraphics[scale=0.35]{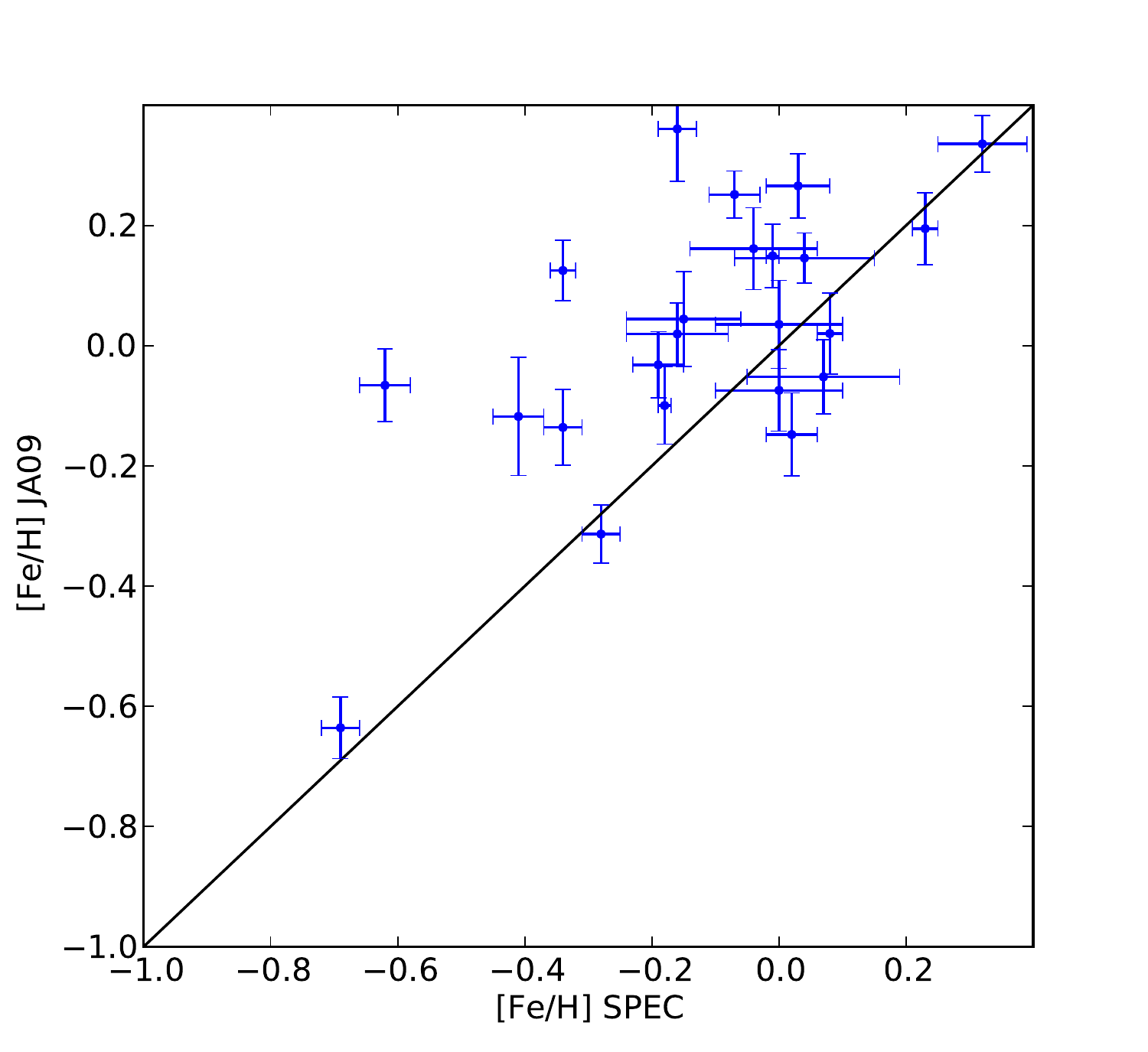}}
\subfigure[SL10 Calibration]{\includegraphics[scale=0.35]{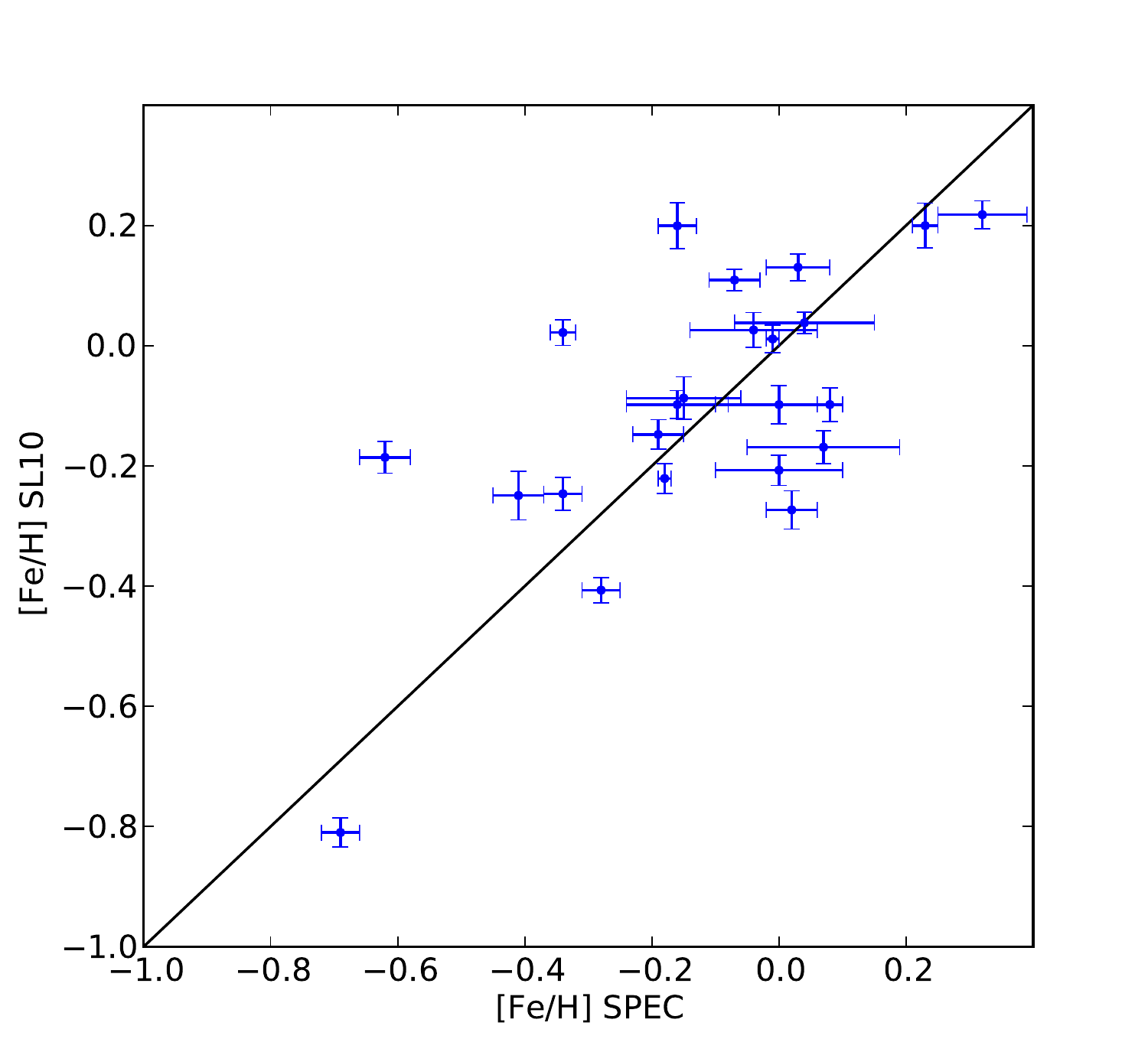}}
\subfigure[This paper]{\includegraphics[scale=0.35]{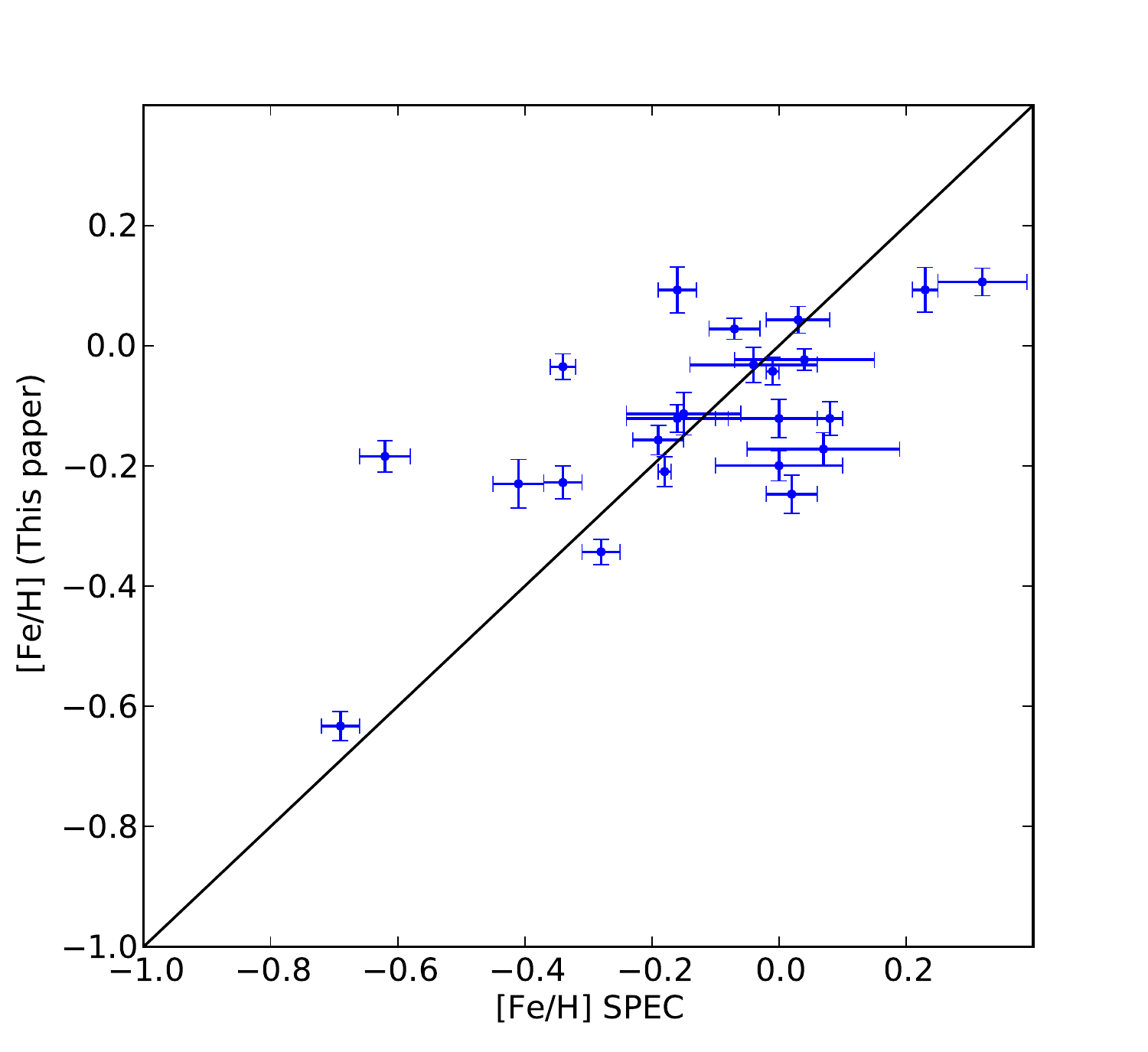}}
\end{center}
\caption{[Fe/H] estimated from the the calibrations versus
spectroscopic metallicity.The blue dots with error bars represent the
data points. The black line depicts a one-to-one relationship.}
\label{fehpanel}
\end{figure*}

In this section we discuss the three photometric metallicity
calibrations in turn, and examine their agreement with our
spectroscopic sample. Figure \ref{fehpanel} plots the [Fe/H] obtained 
from each calibration against the spectroscopic [Fe/H], and it
guides us through that discussion.

\subsection{\citet{Bonfils-2005} calibration}

As recalled in the introduction, B05 first calibrated position
in a $\{(V-K_{s})-M_{K_{s}}\}$ color-magnitude diagram into a useful metallicity
indicator. That calibration is anchored, on the one hand, in
spectroscopic metallicity measurements of early metal-poor M-dwarfs by 
\citet{Woolf-2005}, and on the other hand, in later and more metal-rich 
M~dwarfs which belong in multiple systems for which B05 measured 
the metallicity of a hotter component. The B05 calibration has a
$\sim$0.2 dex dispersion.
Then, they used the calibration to measure the metallicity
distribution of a volume-limited sample of 47 M dwarfs,
which they found to be more metal-poor (by 0.07 
dex\footnote{erroneously quoted as a 0.09 dex difference in
  \citet{Johnson-2009}}) than 1000 FGK stars, with a modest 
significance of 2.6~$\sigma$. 
As mentioned above, \citet{Bonfils-2007} used that calibration 
to compare M dwarfs with and without planets, and found that 
planet hosts are marginally metal-rich.


For our sample, the B05 calibration is offset by $-0.04\pm0.04$ dex
and has a dispersion of 0.20$\pm$0.02~dex. The negative offset is in 
line with SL10 finding (see Section \ref{Schlaufman}) that B05
generally underestimates the true [Fe/H]. Correcting from 
this $-0.04$ offset almost eliminates the metallicity difference between 
local M dwarfs and FGK stars. 

SL10 also report that the B05 calibration has a very poor
$R^{2}_{ap}$, under 0.05, and that their own model explains 
almost an order of magnitude more of the variance of their 
calibration sample. In Sect. 3, we noted, however, that 
$R^2_{ap}$ is a noisy diagnostic for small samples.


In addition to their more commonly used calibration, B05 provide
an alternative formulation for [Fe/H]. That second expression, labeled
B05(2) in Table \ref{stats}, works from the difference between the $V$-
and $K_{s}$-band mass-luminosity relations of \citet{Delfosse-2000}. 
The two B05 formulations perform essentially equally for our
sample, with B05(2) having a marginally higher dispersion. In the 
remainder of this paper we therefore no longer discuss B05(2).

\subsection{\citet{Johnson-2009} calibration}

\citet{Johnson-2009} argue that local M and FGK dwarfs should 
have the same metallicity distribution, and accordingly chose 
to fix their mean M~dwarf metallicity to the value ($-0.05$ dex) 
for a volume-limited sample of FGK dwarfs from 
the \citet{Valenti-2005} sample. They defined a sequence 
representative of average M dwarfs in the $\{(V-K_{s})-M_{K_{s}}\}$ 
color-magnitude diagram, and used the distance to that main sequence 
along $M_{K_{s}}$ as a metallicity diagnostic. They note that the
inhomogeneous calibration sample of B05 is a potential source
of systematics, and consequently chose to calibrate their scale from the 
metallicities of just six metal-rich M dwarfs in multiple systems
with FGK primary components.

JA09 present two observational arguments for fixing the mean M~dwarf 
metallicity. They first measured [Fe/H] for 109 G0-K2 stars
(4900$<$T$_{eff}$$<$5900~K) and found no significant metallicity 
gradient over this temperature range, from which they conclude
that no difference is to be expected for the cooler M dwarfs. We
note, however, that a linear fit to their G0-K2 data set 
([Fe/H]= $9.74\times10^{-5}(T_{\rm eff}-5777)  - 0.04$) 
allows for a wide metallicity range when extrapolated to the 
cooler M dwarfs (2700$<$Teff$<$3750, for M7 to M0 spectral type,
with [Fe/H] $ =  - 0.24$ allowed at the 1~$\sigma$ level 
for an M0 dwarf and significantly lower than the [Fe/H] offset
in B05. More importantly, they measured a large (0.32 dex) 
offset between the B05 metallicities of six metal-rich M dwarfs 
in multiple systems and the spectroscopic metallicities which 
they measured for their primaries. This robustly points to
a systematic offset in the B05 calibration for metal-rich
M~dwarfs, but does not directly probe the rest of the 
(T$_{eff}$, [Fe/H]) space. We do find that the JA09 calibration 
is a good metallicity predictor 
for our sample at high metallicities, where its calibrator was chosen.
With decreasing metallicity, that calibration increasingly 
overestimates the metallicity, however, as previously pointed out by SL10
(see below). Quantitatively, we measure a
$+0.14\pm0.04$~dex offset for our sample and a dispersion of $0.24\pm0.04$.

\subsection{\citet{Schlaufman-2010} calibration}
\label{Schlaufman}
\citet{Schlaufman-2010} improve upon B05 and JA09 in two ways.
They first point out that, for M and FGK dwarfs to share the
same mean metallicity, matched kinematics is as important as 
volume completeness. Since the various kinematic populations 
of our Galaxy have very different mean metallicities, the mean 
metallicity of small samples fluctuates very significantly with 
their small number of stars from the metal-poor populations. To 
overcome this statistical noise, they draw from the Geneva-Copenhagen 
Survey volume-limited sample of F and G stars a subsample that 
kinematically matches the volume limited sample of M dwarfs 
used by JA09. They find a $\simeq-0.14\pm0.06$~dex mean metallicity 
for that sample, $0.09$ dex lower than adopted by JA09. However, they only used 
that sample to verify that the mean metallicity of M dwarfs in the solar neighborhood
is well defined. In the end, the M dwarfs within a sample of binaries with an FGK primary that they used
to fix their calibration are not volume-limited or kinematically-matched, but
their mean metallicity ([Fe/H] = $-0.17\pm0.07$) is statistically indistinguishable from 
the mean metallicity of the volume-limited and kinematically-matched sample. 

Second, they use stellar evolution models to guide their parametrization 
of the color-magnitude space. Using [Fe/H] isocontours for the 
\citet{Baraffe-1998} models, they show that in a \{$(V-K_{s})-M_{K_{s}}$\} 
diagram, changing [Fe/H] affects ($V-K_{s}$) at an essentially constant
$M_{K_{s}}$. The metallicity is therefore best parametrized by ($V-K_{s}$), and
their calibration uses a linear function of the ($V-K_{s}$) distance 
from a nominal sequence in the \{$(V-K_{s})-M_{K_{s}}$\} diagram. They
do not force any specific mean metallicity, but verify {\it a posteriori}
that it matches expectations.

We measure a $0.14\pm0.02$~dex dispersion for the SL10 sample
against their calibration, but that calibration has a significantly 
higher dispersion of $0.19\pm0.03$ for our validation sample. 
That increased dispersion reflects our sample probing a wider
metallicity range than SL10, as verified by computing 
the dispersion of an 18~star subsample that matches the 
metallicity range of the SL10 sample. That dispersion is 
$0.15\pm0.03$~dex, and indistinguishable from $0.14\pm0.02$~dex
for the SL10 sample. The increased dispersion for a wider
metallicity range suggests that a linear function of ($V-K_{s}$)
does not fully describe metallicity.
We also measure an offset of $0.02\pm0.04$~dex. Offset and rms both improve
over either of the B05 and JA10 calibrations.

\subsection{Refining the \citet{Schlaufman-2010} calibration}


We produced updated coefficients for the SL10 prescription, using
the $RMS_{p}$ free parameter $p = 2$ (see Sect. \ref{test}). The
expression for the new metallicity calibration is

\begin{eqnarray}
[Fe/H] = 0.57 \Delta (V-K_{s}) - 0.17, \\
\indent \Delta(V-K_{s}) = (V-K_{s})_{obs}-(V-K_{s})_{iso}, \nonumber
\label {sl10new}
\end{eqnarray}
where $(V-K_{s})_{obs}$ is the observed $V-K_{s}$ color 
and $(V-K_{s})_{iso}$ is a fifth-order polynomial function of 
$M_{K_{s}}$ that describes the mean main sequence of the solar 
neighborhood from the \citet{Valenti-2005} catalog. This expression is adopted
from \citet{Schlaufman-2010},  who adapted an $M_{K_{s}}$ vs
$(V-K_{s})$ formula from \citet{Johnson-2009}. 

Table \ref{stats} shows limited differences between this new fit 
and the original SL10 calibration. The dispersion of the new fit 
is tighter by just 0.02 dex ($0.17\pm0.03$~dex instead of
$0.19\pm0.03$), and the offset is now $0.00\pm0.04$, as expected. 
The R$^{2}_{ap}$ value is similar ($0.43\pm0.23$ vs $0.41\pm0.29$) 
and uncertain. Readjusting the coefficients therefore produces
a marginal improvement at best.

The dispersion, in all panels of Fig.~\ref{fehcalib}, is well
above the measurement uncertainties. Those therefore contribute
negligibly to the overall dispersion, which must be dominated
by other sources.

As can be seen in Fig. \ref{fehpanel}, B05 or B05(2) tend to
underestimate [Fe/H], while the JA09 calibration clearly 
overestimates [Fe/H] except at the highest metallicities.

\section {Summary}
\label{discussion}

We have assembled a sample of M dwarf companions to hotter FGK
stars, where the system has an accurate parallax and the M~dwarf
component has accurate $V$ and $K_{s}$-band photometry. Using the metallicities
of the primaries, newly measured or retrieved from the literature, 
and the assumption that the two components have identical initial 
compositions, we compared the dispersions of the \citet{Bonfils-2005}, 
\citet{Johnson-2009}, and \citet{Schlaufman-2010} photometric 
metallicity calibrations. We find that the \citet{Schlaufman-2010} 
scale, which is intermediate between \citet{Bonfils-2005} and
\citet{Johnson-2009}, has the lowest dispersion. We slightly
refine that relation, by readjusting its coefficients on our 
sample. 

We find that our tight selection of binaries with accurate 
parallaxes and photometry sample has insignificantly reduced 
the dispersion of the measurements around the calibration 
compared to looser criteria. This suggests that the dispersion, 
hence the random errors of the calibration, is not
defined by measurement uncertainties but instead reflects 
intrinsic astrophysical dispersion. Nonlinearities in the
metallicity dependence of the $V-K_{s}$ color are likely
to contribute, as suggested both by atmospheric models
(Allard, private communication) and by the increased
dispersion that we measure over a wider metallicity range.
They are, however, unlikely to be the sole explanation,
since we see dispersion even in narrow areas of the
color-magnitude diagram. Stellar evolution cannot significantly
contribute, since early-M dwarfs evolve rapidly to the main
sequence and they remain there for much longer than a Hubble time,
but rotation and magnetic activity could play a role.
Unless, or until, we develop a quantitative understanding of
this astrophysical dispersion, the photometric calibration 
approach may therefore have reached an intrinsic limit. 
Those calibrations also have the very practical inconvenience 
of needing an accurate parallax. This limits their use to the 
close solar neighborhood, at least until the GAIA catalog becomes 
available in a decade.

Alternative probes of the metallicities of M dwarfs are therefore
obviously desirable. One obvious avenue is to work from higher 
spectral resolution information and to identify spectral elements
that are most sensitive to metallicity and others that are 
most sensitive to effective temperature. We are pursuing this
approach at visible wavelengths (Neves et al. in prep.), as
do \citet[see Appendix \ref{rojas}]{Rojas-Ayala-2010} in the near 
infrared, with encouraging results in both cases.

\begin{acknowledgements}
%
We would like to thank Luca Casagrande for kindly providing the
metallicities calculated from his calibration. We also would like to thank 
Barbara Rojas-Ayala for finding an error in the text regarding the absolute magnitudes. We acknowledge the
support by the European Research Council/European Community under the
FP7 through Starting Grant agreement number 239953. NCS also
acknowledges the support from Funda\c{c}\~ao para a Ci\^encia e a
Tecnologia (FCT) through program Ci\^encia\,2007 funded by FCT/MCTES
(Portugal) and POPH/FSE (EC), and in the form of grant reference
PTDC/CTE-AST/098528/2008. VN would also like to acknowledge the
support from the FCT in the form of the fellowship SFRH/BD/60688/2009.

\end{acknowledgements}

\bibliographystyle{aa}
\bibliography{mylib.bib}

\appendix

\section{Other methods}

\subsection{Calibration of \citet{Casagrande-2008}}
\label{casagrande}
In Sect. \ref{latest} we described the photometric
metallicity calibrations in detail. \citet{Casagrande-2008} devised a 
completely different technique, based on their previous study of 
FGK stars using the infrared flux method \citep{Casagrande-2006}, 
to  determine effective temperatures and metallicities. The infrared 
flux method uses multiple photometry bands to derive effective
temperatures, bolometric luminosities, and angular diameters.
The basic idea of IRFM \citep{Blackwell-1977} is to compare the
ratio between the bolometric flux and the infrared monochromatic flux,
both measured on Earth, to the ratio between the surface bolometric
flux ($\propto\sigma Teff^{4}$) and the surface infrared monochromatic
flux for a model of the star. To adapt this method to M dwarfs,  
\citet{Casagrande-2008} added optical bands, creating the so-called 
MOITE, Multiple Optical and Infrared TEchnique. This method provides 
sensitive indicators of both temperature and metallicity. The 
proposed effective temperature scale extends down to 2100-2200 K, 
into the L-dwarf limit, and is supported by interferometric angular 
diameters above $\sim$ 3000K. \citet{Casagrande-2008} obtain
metallicities by computing the effective temperature of the
star for each color band $(V(RI)_{c}JHK_{s})$ for several 
trial metallicities, between $-$2.1 and 0.4 in 0.1~dex steps,
and by selecting the metallicity that minimizes the scatter among 
the six trial effective temperatures. \citet{Casagrande-2008} 
estimate that their total metallicity uncertainty is 0.2 to 0.3~dex.

\begin{table}[]
\caption{Metallicity values from spectroscopy and obtained using the method of \citet{Casagrande-2008} (C08 in this table).}
\label{tablefehc08}
\begin{center}
\begin{tabular}{r r r r}

\hline
\hline

Primary & Secondary & \multicolumn{2}{c}{[Fe/H] [dex]} \\
 & & Spectroscopic & C08 \\
\hline

Gl53.1A  & Gl53.1B  & 0.07 & -0.07 \\
Gl56.3A  & Gl56.3B  & 0.00 & -0.21 \\
Gl81.1A  & Gl81.1B  & 0.08 & -0.08 \\
Gl100A  & Gl100C  & -0.28 & -0.10 \\
Gl105A  & Gl105B  & -0.19 & -0.30 \\
Gl140.1A  & Gl140.1B  & -0.41 & -0.30 \\
Gl157A  & Gl157B  & -0.16 & -0.10 \\
Gl173.1A  & Gl173.1B  & -0.34 & -0.20 \\
Gl211  & Gl212  & 0.04 & -0.21 \\
Gl231.1A  & Gl231.1B  & -0.01 & -0.28 \\
Gl250A  & Gl250B  & -0.15 & - \\
Gl297.2A  & Gl297.2B  & 0.03 & 0.00 \\
Gl324A  & Gl324B  & 0.32 & -0.20 \\
Gl559A  & Gl551  & 0.23 & - \\
Gl611A  & Gl611B  & -0.69 & -0.40 \\
Gl653  & Gl654  & -0.62 & -0.30 \\
Gl666A  & Gl666B  & -0.34 & - \\
Gl783.2A  & Gl783.2B  & -0.16 & -0.30 \\
Gl797A  & Gl797B  & -0.07 & -0.90 \\
GJ3091A  & GJ3092B  & 0.02 & -0.30 \\
GJ3194A  & GJ3195B  & 0.00 & -0.60 \\
GJ3627A  & GJ3628B  & -0.04 & -0.20 \\
NLTT34353  & NLTT34357  & -0.18 & 0.19 \\

\hline
\end {tabular}
\end{center}
\end{table}


%
%

\begin{figure}[h]
\begin{center}
\includegraphics[scale=0.5]{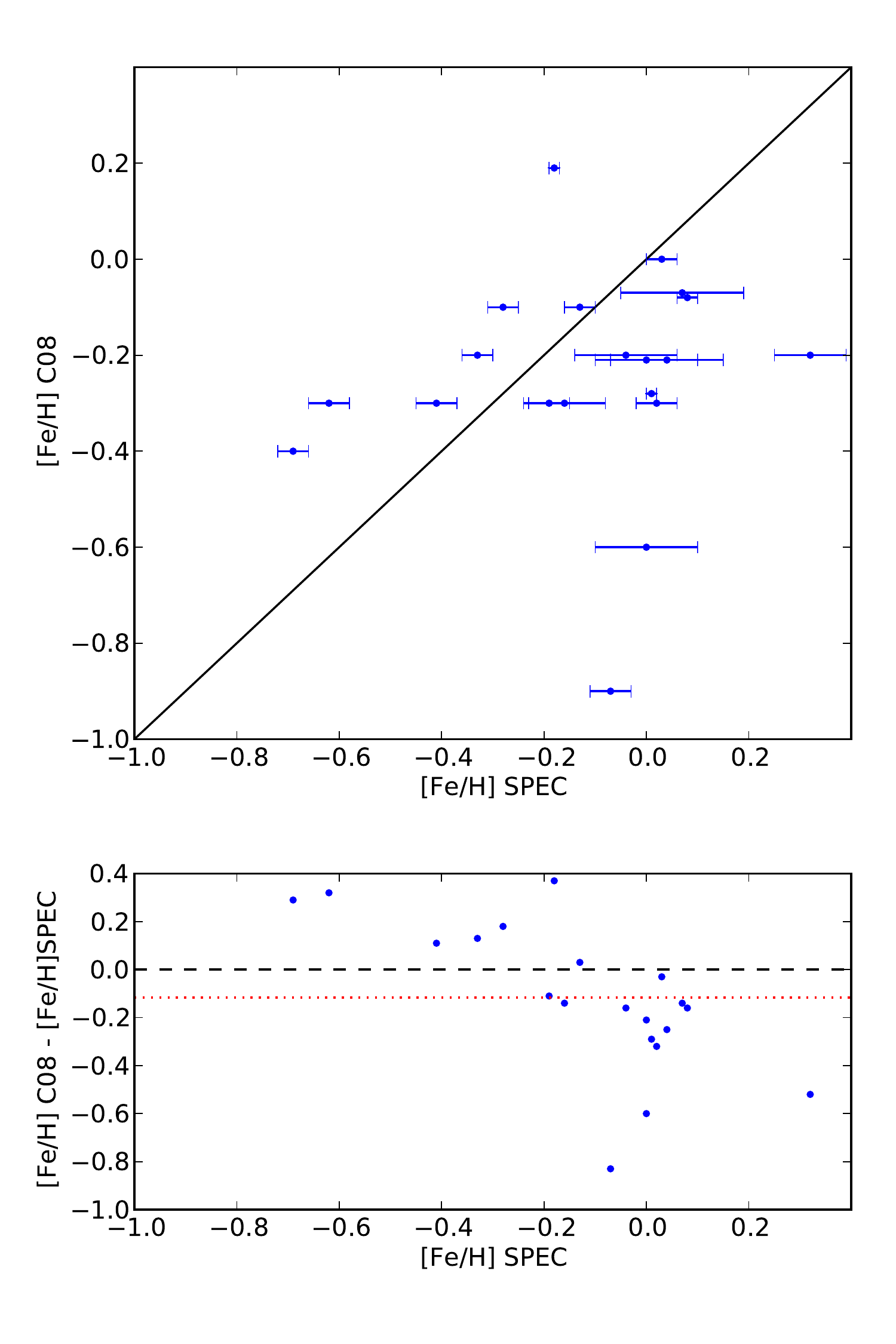}
\end{center}
\caption{[Fe/H] obtained with the \citet{Casagrande-2008} method
  versus the spectroscopic metallicity. The blue dots with error bars
  represent the data points. The black line depicts a one-to-one
  relationship. The metallicity difference between the values of the
  calibrations and the spectroscopic measurements is shown below each
  [Fe/H]-[Fe/H] plot. The black dashed line is the zero point of the
  difference, and the red dotted line represents the average of the
  metallicity difference.  }
\label{fehfehc08}
\end{figure}

The MOITE method does not reduce into a closed form that 
can be readily applied by others, but Luca Casagrande kindly 
computed MOITE [Fe/H] values for our sample (Table~\ref{tablefehc08}). 
We evaluated the calibration in the same manner as in Sect. \ref{test} and obtained 
a value of $-0.11\pm0.07$ dex for the offset, $0.32\pm0.06$ dex for the rms, $0.10\pm0.04$ dex
for the RMS$_{P}$, and $-1.09\pm1.45$ for the R$^{2}_{ap}$. 
From these values and from Fig.~\ref{fehfehc08}, we can observe that 
the \citet{Casagrande-2008} calibration has a higher rms and RMS$_{p}$ and a poorer R$^{2}_{ap}$ 
than the three photometric calibrations, consistently with the high
metallicity uncertainty referred by \citet{Casagrande-2008}. The
negative R$^{2}_{ap}$ value formally means that this model increases 
the variance over a constant metallicity model, but as usual
R$^{2}_{ap}$ is a noisy diagnostic.

\subsection{\citet{Rojas-Ayala-2010} calibration}
\label{rojas}

\citet{Rojas-Ayala-2010} have recently published a novel and potentially
very precise technique for measuring M dwarf metallicities. Their
technique is based on spectral indices measured from 
moderate-dispersion ($R \sim 2700$) $K$-band spectra, and it needs 
neither a $V$ magnitude nor a parallax, allowing measurement of 
fainter (or/and farther) stars. They
analyzed 17 M dwarf secondaries with an FGK primary, which also served 
as metallicity calibrators, and measured the equivalent widths of the
NaI doublet (2.206 and 2.209 $\mu m$), and the CaI triplet (2.261,
2.263 and 2.265 $\mu m$). With these measurements and a water
absorption spectral index sensitive to stellar temperatures, they
constructed a metallicity scale with an adjusted multiple correlation
coefficient greater than the one of \citet{Schlaufman-2010}
($R^{2}_{ap} = 0.63$), and also with a tighter RMS$_{p}$ of 0.02 when
compared to other studies (0.05, 0.04, and 0.02 for
\citeauthor{Bonfils-2005} \citeyear{Bonfils-2005}, \citeauthor{Johnson-2009} \citeyear{Johnson-2009}, and
\citeauthor{Schlaufman-2010} \citeyear{Schlaufman-2010} respectively). The metallicity calibration is 
valid over -0.5 to +0.5 dex, with an estimated
uncertainty of $\pm$0.15 dex.


A test of the \citet{Rojas-Ayala-2010} calibration for our full sample
would be very interesting, but is not currently possible for lack of
near-infrared spectra for most of the stars. Seven of our stars,
however, have their metallicities measured in \citet{Rojas-Ayala-2010} 
(Gl 212, Gl 231.1B, Gl 250B, Gl 324B, Gl611B, Gl783.2B, and Gl 797B
with predicted [Fe/H] of $0.09, -0.05, -0.04, 0.30, -0.49, -0.19$, and $
 -0.06$ dex, respectively). We find a dispersion of only 0.08~dex and an
offset of 0.04~dex offset between our spectroscopic measurements 
of the primaries and the \citet{Rojas-Ayala-2010} metallicities of the
secondaries. These numbers are extremely encouraging, but still have little 
statistical significance. They will need to be bolstered by testing
against a larger sample and over a wider range of both metallicity and
effective temperature.

\end{document}